\documentclass{aa}

\newcommand{\pc}[1]{\ensuremath{\left(#1\right)}}

\usepackage{pifont}% http://ctan.org/pkg/pifont
\usepackage{tikz}

\usepackage{bm}
\usepackage{soul}
\usepackage{multirow}
\usepackage{longtable,tabu}
\newcommand{\be}{\begin{equation}}
\newcommand{\ee}{\end{equation}}
\newcommand{\bea}{\begin{eqnarray}}
\newcommand{\eea}{\end{eqnarray}}

\usepackage{amsmath}

\usepackage{url}

\usepackage{xcolor}

\definecolor{color1}{HTML}{440154}
\definecolor{color2}{HTML}{481568}
\definecolor{color3}{HTML}{482677}

\definecolor{color18}{HTML}{B8DE29}
\definecolor{color19}{HTML}{DCE318}
\definecolor{color20}{HTML}{FDE725}

\graphicspath{{./}{figures/}}
\begin{document}

\title{Light clusters as a possible source of crustal impurities: a quasi-particle approach}

\author{Helena Pais$^1$\thanks{hpais@uc.pt} \and Hoa Dinh-Thi$^2$ \and Anthea F. Fantina$^3$ \and Francesca Gulminelli$^4$ \and Constan\c ca Provid\^encia$^1$}

\institute{$^1$CFisUC, University of Coimbra, P-3004-516 Coimbra, Portugal.\\
$^2$Department of Physics and Astronomy - MS 108, Rice University, 6100 Main Street, Houston, TX 77251-1892, USA. \\
$^3$Grand Accélérateur National d’Ions Lourds (GANIL), CEA/DRF - CNRS/IN2P3, Boulevard Henri Becquerel, 14076 Caen, France. \\
$^4$Université Normandie, ENSICAEN, LPC-Caen, 14000 Caen, France.} 

\date{Received xxx Accepted xxx}

%---

 \abstract
  {The presence of impurities in the neutron star crust is known to affect in an important way the thermal and electrical conductivity of the star.} 
{In this work, we explore the possibility that such impurities might arise from the simultaneous presence of heavy ions together with Hydrogen and Helium isotopes formed during the cooling process of the star.}   
{We consider an equilibrium population of such light particles at temperatures close to the crystallization of the crust within an effective quasi-particle approach 
including in-medium binding energy shifts, and using different versions of the relativistic mean field approach for the crustal modeling. Thermal effects are consistently included also in the dominant ion species present in each crustal layer described in the compressible liquid drop approximation.}
{We find that the impurity factor associated to light clusters is comparatively very small and can be neglected in transport calculations, even if a strong model dependence is observed.  }
   {}


\authorrunning{Pais et al.}

\maketitle

\section{Introduction} \label{sec:introduction}

Light nuclear clusters such as Hydrogen or Helium isotopes are expected to be produced in hot and dense stellar matter, and their abundance is known to affect the dynamical properties of core-collapse supernovae \citep{Arcones:2008kv,Sumiyoshi:2008qv,Furusawa:2013tta,Furusawa:2016tdj,Fischer:2013eka,Fischer:2020krf} and binary neutron star (NS) mergers \citep{Bauswein:2013yna,Rosswog:2015nja,Radice:2018pdn,Psaltis:2023jvk}. Even if the weak rates directly involving light clusters are low \citep{Fischer:2020krf}, their abundance is in competition with the one of heavy nuclei \citep{Pais:2019shp}, which indirectly affects the neutrino dynamics, and the size and proton fraction of the nucleosynthesis seeds \citep{Nedora:2020hxc}. 

The possible relevance of light clusters in NS physics is much less addressed in the literature.
In this paper, we explore the possibility that such light composites might play the role of impurities in the crustal lattice.
Formed from core-collapse supernova explosions, NSs are indeed born hot, with temperatures exceeding several MeV \citep{Haensel:2007yy}. In these conditions, the crust of the  proto-NS is a liquid plasma
composed of different nuclear species including light clusters in a background of electrons and nucleons.
It is generally assumed that, as the NS crust cools down, this multi-component plasma remains in full thermodynamic equilibrium until the crust crystallizes, and that the composition at the crystallization temperature coincides with the ground-state composition, namely a sequence of pure layers, each consisting of a one-component ideal Coulomb crystal. 
In this situation, the only nuclear species relevant for the description of a NS are heavy ions with atomic numbers of the order $Z\approx 20-40$.

However, the crustal lattice is not expected to be strictly perfect, and can present different defects that are collectively known as impurities and quantified via an ``impurity parameter'' that measures the importance of charge fluctuations at each depth of the crust. 
In the magneto-thermal simulations of NS evolution, this (possibly density-dependent) impurity parameter is typically adjusted on observational cooling data \citep{Vigano:2013lea}.
A high value for the impurity parameter lowers the electrical and thermal conductivity, and seems to be needed for a better description of different astrophysical observations concerning isolated X-ray pulsars and late time cooling in binaries \citep{Pons:2013nea,Newton:2013,Deibel:2016vbc,Hambaryan:2017wvm,Tan:2018rhg}.

If the presence of highly resistive layers due to the presence of impurities seems well settled, its physical origin and the actual value of the impurity parameter are less clear. 
Classical molecular dynamics simulations suggest that  the innermost part of the inner crust is composed of  disordered and amorphous nuclear matter, with coexistence of different exotic complex shapes \citep{Schneider:2014lia,Horowitz:2014xca,Caplan:2020ewl,Newton:2021vyd}. The onset of this pasta mantle is, however, model dependent \citep{Thi:2021,Shchechilin:2023erz} and the most extended part of the inner crust is believed to be composed of standard spherical nuclei \citep{Shchechilin:2024kjv}. 
In the spherical regime, if the one-component plasma approximation is released, it was suggested that different ion species  might persist as impurities in the whole catalyzed crust, due to the non-adiabaticity of the cooling dynamics \citep{Goriely:2011er,Fantina:2019lbd,Carreau:2020gth,Potekhin:2020rcs,DinhThi:2023ioy}. 
In particular, the authors of Refs.~\cite{Fantina:2019lbd,DinhThi:2023ioy} computed the impurity factor self-consistently within a multi-component approach at different temperatures close to the crystallization temperature. 
They found that high values of the impurity parameter between $Q_{\rm imp}\approx 5$ and $Q_{\rm imp}\approx 100$ can be obtained in the inner crust, even in the density region where deformed pasta structures are not expected, due to the simultaneous presence at finite temperature of light ($Z\approx 2-3$) and heavy ($Z\approx 20-30$) ions. 
In that study, a single energy functional was employed, and the cluster free energy was computed with the compressible liquid drop (CLD) model, which gives a poor description of the lightest nuclei. It is therefore interesting to study the model dependence of those results, and to explore a more accurate and microscopic description of the light cluster free energies in a dense nuclear medium.

From the study of Refs.~\cite{Fantina:2019lbd,DinhThi:2023ioy} we expect that important charge fluctuations can only occur if the charge distribution has a multi-modal structure, with clusters of very different sizes appearing with comparable probability. 
We therefore consider a simplified theoretical setting where the one-component plasma approximation is kept for the dominant ion species, but Hydrogen and Helium isotopes are additionally added as explicit degrees of freedom, as it was proposed in previous works employing the relativistic mean-field (RMF) approximation for the nuclear interaction \citep{Avancini:2012bj,Ferreira:2012ha}. The advantage of this description is that in-medium modifications of the light cluster energies in the medium are easily introduced as a modification of the couplings of nucleons bound in clusters with the effective mesonic fields \citep{Pais:2018vac}. This effective coupling modification was extensively calibrated on ab-initio calculations \citep{Ropke:2014fia,Ropke:2020,Ren:2023ued} and experimental constraints \citep{Qin:2012,Bougault:2019djd} in particular for the FSU functional \citep{Pais:2019jst,Custodio:2020bjz,Custodio:2024ost}.
Within this formalism, fluctuations of the dominant ion species in the different Wigner-Seitz cells are neglected, as well as the possible presence of exotic neutron-rich clusters \citep{Pais:2019shp}. 
The resulting $Q_{\rm imp}$ values will, therefore, represent a lower limit to the crustal impurities.

The paper is organized as follows: in Sect.~\ref{sec:method} we introduce the formalism used in this work; specifically, we describe in Sect.~\ref{sec:matter} the treatment of homogeneous matter, in Sect.~\ref{sec:CLDM} that of the (heavier) clusters through the CLD model, and in Sect.~\ref{sec:light-clusters} the treatment of light clusters. We present our numerical results in Sect.~\ref{sec:results}, and finally we draw our conclusions in Sect.~\ref{sec:conclusions}. In this work, we consider natural units, where $\hbar=c=k_B=1$.

\section{Formalism \label{sec:method}}

Each layer of the inner crust of the NS is described as a collection of identical neutral cells constituted of a single spherical nucleus, and a uniform distribution of dripped neutrons and point-like light clusters, all interacting with and through effective meson fields.
The nucleus is treated as a CLD with bulk, surface, and Coulomb terms, in the same spirit as in Ref.~\cite{Pais15}. Since we are considering the liquid phase of the crust, translational degrees of freedom are also associated to the nuclear droplet.

In the following subsections we describe in further details the theoretical treatment of homogeneous matter (Sect.~\ref{sec:matter}), that is used for the bulk part of the droplet energy as well as for the dripped neutrons, the finite-size and translational terms terms associated to the nuclear droplet within a CLD model (Sect.~\ref{sec:CLDM}), and the treatment of the light clusters (Sect.~\ref{sec:light-clusters}).

\subsection{Homogeneous matter}
\label{sec:matter}

Our formalism is based on the non-linear Walecka model in the mean-field approximation \citep{Furnstahl:1995zb,Furnstahl:1996wv}, where we consider an infinite homogeneous  system of nucleons with mass $M$ (protons $p$ and neutrons $n$), interacting with and through three meson fields: an isoscalar-scalar field $\phi$ with mass $m_s$, an
isoscalar-vector field $V^{\mu}$ with mass $m_v$ and an
isovector-vector field $\mathbf b^{\mu}$ with mass $m_\rho$, for the isospin dependence.  Protons and electrons interact through the
electromagnetic field $A^{\mu}$. The Lagrangian density is given as:

\begin{equation} \label{eq:lagrangian}
    \mathcal{L}=\sum_{j=p,n}{\mathcal{L}_{j}} +\mathcal{L}_{e } 
    + \mathcal{L}_{{\sigma }}+ \mathcal{L}_{{\omega }} + 
\mathcal{L}_{{\rho }}+{\cal L}_{\omega \rho}+{\mathcal{L}}_{\gamma } .%,
\end{equation}
The nucleon and electron terms read, respectively,
\begin{eqnarray}
\mathcal{L}_{j}&=&\bar{\psi}_{j}\left[ \gamma _{\mu }iD^{\mu }-M^{*}\right]
\psi _{j}  \label{lagnucl}, \\
iD^{\mu }&=&i\partial ^{\mu }-g_vV^{\mu }-\frac{g_{\rho }}{2}{\boldsymbol{\tau}}%
\cdot \boldsymbol{b}^{\mu } - e \frac{1+\tau_3}{2}A^{\mu}, \nonumber \\  \\
\mathcal{L}_e&=&\bar \psi_e\left[\gamma_\mu\left(i\partial^{\mu} + e A^{\mu}\right)
-m_e\right]\psi_e \, . \label{lage}
\end{eqnarray}
and the nucleon Dirac effective mass is given as
\begin{eqnarray}
M^{*} &=&M-g_s\phi \, .
\end{eqnarray}
The meson and electromagnetic fields Lagrangian densities are given by:
\begin{eqnarray*}
\mathcal{L}_{{\sigma }} &=&\frac{1}{2}\left( \partial _{\mu }\phi \partial %
^{\mu }\phi -m_{s}^{2}\phi ^{2} -\frac{1}{3}\kappa \phi ^{3}-\frac{1}{12}%
\lambda \phi ^{4} \right) \\ 
\mathcal{L}_{{\omega }} &=&-\frac{1}{4}\Omega _{\mu \nu }\Omega ^{\mu \nu }+ \frac{1}{2}m_v^{2}V_{\mu }V^{\mu }+\frac{1}{4!}\xi g_{v}^{4}(V_{\mu}V^{\mu })^{2} \\
\mathcal{L}_{{\rho }} &=&-\frac{1}{4} \mathbf{B}_{\mu \nu }\cdot \mathbf{B}^{\mu\nu }+ \frac{1}{2} m_{\rho }^{2}\boldsymbol{b}_{\mu }\cdot \boldsymbol{b}^{\mu }\\
\mathcal{L}_{{\gamma }} &=&-\frac{1}{4}F _{\mu \nu }F^{\mu
  \nu }\\ 
{\cal L}_{\omega \rho}&=&\Lambda_{v} g_v^2 g_\rho^2 \mathbf
b_{\mu}\cdot \mathbf b^{\mu}\, V_{\mu}V^{\mu} \, ,
\end{eqnarray*}
where $\Omega _{\mu \nu }=\partial _{\mu }V_{\nu }-\partial
_{\nu }V_{\mu }$, $\mathbf{B}_{\mu \nu }=\partial _{\mu
}\boldsymbol{b}_{\nu }-\partial _{\nu }\boldsymbol{b}_{\mu
}-g_{\rho }(\boldsymbol{b}_{\mu }\times
\boldsymbol{b}_{\nu })$ and $F_{\mu \nu }=\partial _{\mu
}A_{\nu }-\partial _{\nu }A_{\mu }$. The nonlinear ${\cal L}_{\omega \rho}$ term affects the density dependence of the symmetry energy.
The models comprise the following parameters: three coupling constants $g_s$, $g_v$ and $g_{\rho}$ of the mesons to the nucleons, the mixing $\omega\rho$ coupling, $\Lambda_{v}$, the bare nucleon mass $M$, the electron mass $m_e$, the masses of the mesons, the electromagnetic coupling constant $e=\sqrt{4 \pi/137}$ and the self-interacting coupling constants $\kappa$, $\lambda$, and $\xi$. 
In this Lagrangian density, $\boldsymbol \tau$ are the Pauli matrices.

Concerning the choice of the nuclear interaction, 
we use two families of models, non-linear (NL) models, FSU \citep{Todd-Rutel:2005yzo}, FSU2R \citep{Negreiros:2018cho}, NL3$\omega\rho$ \citep{Horowitz:2000xj,Pais:2016xiu}, and NLset21 \citep{Malik:2024nva}, whose couplings of the mesons to the nucleons are fixed, and the meson terms included in the Lagrangian density are up to quartic order, as defined above, and density-dependent (DD) models, namely TW \citep{Typel:1999yq}, DD2 \citep{Typel:2009sy} and DDME2 \citep{Lalazissis:2005}, whose couplings of the mesons to the nucleons vary with the density. In this case, the self-interaction couplings $\kappa,\lambda$ and the mixing $\omega\rho$ coupling $\Lambda_v$ are zero, and the $g_s$, $g_v$ and $g_{\rho}$ couplings are given by the density-dependent functions \citep{Typel:1999yq,Typel:2009sy,Lalazissis:2005}
\begin{eqnarray}
g_i&=&g_{i_0} f_i(x) \, , i=s,v \ , \\
f_i(x)&=&a_i\frac{1+b_i(x+d_i)^2}{1+c_i(x+d_i)^2} \ , \\
g_{\rho}&=&g_{\rho_0} \exp(-a_{\rho}(x-1)) \ ,
\end{eqnarray}
with $x=\rho/\rho_0$.
The reader can refer to Refs.~\cite{Typel:1999yq,Typel:2009sy,Lalazissis:2005} for the values of the parameters $a_i,b_i,c_i,d_i$ for each DD model.

DDME2  was fitted to reproduce the properties of symmetric and asymmetric matter, binding energies and neutron and charge radii of stable nuclei. Moreover, this model is able to reproduce two-solar-mass NSs. 
DD2 also reproduces finite nuclei properties and is based on the TW model, the first density-dependent model that appeared in the literature. 
FSU, on the other hand, is not able to produce two-solar-mass NSs, but it describes well the properties of nuclear matter at and below saturation density \citep{Grill:2014aea}, which is the range of densities considered in this work. 
However, FSU2R was constructed in order to correct the two-solar-mass problem, and differs from FSU in the isovector sector.  
NLset21 is a recent model chosen from a set of NL models constructed using Bayesian techniques \citep{Malik:2022jqc} that fulfill astrophysical, chiral effective-field-theory (EFT) neutron-matter calculations, and also reproduce well the symmetric matter properties at saturation. 
NL3$\omega\rho$ is a model from the NL3 family, i.e. it has the same isoscalar properties as NL3, but differs in the isovector properties. Since NL3 has a very large slope of the symmetry energy, $L_{\rm sym,0}$ = 118~MeV, the NL3$\omega\rho$ model is constructed by adding a term mixing the $\omega$ and $\rho$ mesons in order to lower the value of 
$L_{\rm sym, 0}$: the mixing term is varied and the corresponding isovector coupling is calculated so that the symmetry energy at $\rho=0.1$ fm$^{-3}$ has the same value as in NL3.

The  equilibrium state of asymmetric nuclear matter is characterized by the distribution functions, $f_{0k\pm}$, of particles ($+$) and antiparticles ($-$) $k=p,n,e$, given by: 
\begin{equation}
f_{0 j \pm}= \frac{1}{1+e^{(\epsilon_{0 j} \mp \nu_j)/T}}, \quad j=p,n
\end{equation}
with
\begin{equation}
\epsilon_{0 j}=\sqrt{p^2+{M^*}^2},  \quad \nu_j=\mu_j - g_v V_0^{(0)}  - \frac{g_\rho}{2}\, \tau_j b_0^{(0)}  \label{chempot}
\end{equation}
and
\begin{equation}
f_{0 e \pm}= \frac{1}{1+e^{(\epsilon_{0 e} \mp \mu_e)/T}}, 
\end{equation}
\begin{equation}
\epsilon_{0e}=\sqrt{p^2+m_e^2},
\end{equation}
where $\mu_k$ is the chemical potential of particle $k=p,n,e$.
The net particles densities are readily calculated from the distribution functions as
\begin{equation} \label{eq:density}
     \rho_j=\frac{1}{\pi^2} \int dp\,p^2 \left(f_{0j+}+f_{0j-}\right) \, .
\end{equation}
In the mean-field approximation, the thermodynamic quantities of interest are given in terms of the meson fields, which are replaced by their constant expectation values (denoted by a subscript `0'). 

For homogeneous neutral nuclear matter, the energy density, the entropy density, the free energy density, and the pressure are given, respectively, by:
\begin{eqnarray}
\mathcal{E}&=&\frac{1}{\pi^2}\sum_{j=p,n}\int dp\,p^2\epsilon_{0j}\left(f_{0j+}+f_{0j-}\right) \nonumber \\
&+&\frac{m_v^2}{2}V_0^2+\frac{\xi g_v^4}{8}V_0^4+\frac{m_\rho^2}{2}b_0^2+\frac{m_s^2}{2}\phi_0^2 \nonumber \\
&+&\frac{k}{6}\phi_0^3+\frac{\lambda}{24}\phi_0^4+3\Lambda_v g_\rho^2g_v^2V_0^2b_0^2 \ , \label{energy} \\
\mathcal{S}&=&-\frac{1}{\pi^2}\sum_{j=p,n}\int dp\,p^2 \left[f_{0j+}\ln f_{0j+} \right. \nonumber \\
&+&(1-f_{0j+})\ln(1-f_{0j+})+ f_{0j-}\ln f_{0j-} \nonumber \\
&+&\left.(1-f_{0j-})\ln(1-f_{0j-})\right] \ , \label{entropy}\\
\mathcal{F}&=&\mathcal{E}-TS \ ,\label{free} \\
P&=&\frac{1}{3\pi^2}\sum_{j=p,n}\int dp\frac{p^4}{\epsilon_{0j}}\left(f_{0j+}+f_{0j-}\right) \nonumber \\
&+&\frac{m_v^2}{2}V_0^2+\frac{\xi g_v^4}{24}V_0^4+\frac{m_\rho^2}{2}b_0^2-\frac{m_s^2}{2}\phi_0^2 \nonumber \\
&-&\frac{k}{6}\phi_0^3-\frac{\lambda}{24}\phi_0^4+\Lambda_v g_\rho^2g_v^2V_0^2b_0^2 \ . \label{pressure} 
\end{eqnarray}
For the DD models, there is an extra term in the pressure $P\rightarrow P+ \rho\Sigma_0$ with 
\begin{eqnarray}
\rho\Sigma_0=\rho[g'_v V_0\rho+g'_{\rho}b_0(\rho_p-\rho_n)-g'_s\phi_0\rho_s] \, 
\end{eqnarray}
with $g'_i$ the derivative of the couplings with respect to the density. 
This rearrangement term, $\Sigma_0$, appears in the vector self-energy due to the density dependence of the couplings.

For the electrons, we have
\begin{eqnarray}
\mathcal{E}_e&=&\frac{1}{\pi^2}\int dp\, p^2  \epsilon_{0e}\left(f_{0e+}+f_{0e-}\right) \ , \label{energy_e}\\
\mathcal{S}_e&=&-\frac{1}{\pi^2}\int dp\,p^2 \left[f_{0e+}\ln f_{0e+}+(1-f_{0e+})\ln(1-f_{0e+})\right. \nonumber \\
&+&\left. f_{0e-}\ln f_{0e-}+(1-f_{0e-})\ln(1-f_{0e-})\right] \ , \\
\mathcal{F}_e&=&\mathcal{E}_e-TS_e \ , \label{free_e}\\
P_e&=&\frac{1}{3\pi^2}\int dp \frac{p^4}{\epsilon_{0e}}\left(f_{0e+}+f_{0e-}\right) \ . \label{pressure_e}
\end{eqnarray}

\subsection{Inhomogeneous matter in the compressible liquid drop model}
\label{sec:CLDM}

The simplest treatment of a nucleus immersed in a sea of dripped nucleons is to divide each cell of volume $V_W$ into two separated regions of constant density, such that $V_W=V^I+V^{II}$, the higher density region (I) representing the nucleus, and the lower density one (II) representing the background nucleon gas. 
The baryonic number and charge of the nucleus are therefore given by:
 \begin{eqnarray}
A_{\rm heavy}&=&V^I \rho^I \ , \label{eq:Aheavy} \\
Z_{\rm heavy}&=&V^I \rho^I y_p^I \ . \label{eq:Zheavy}
\end{eqnarray}
In this picture, the interface between the two regions is sharp, and all the thermodynamical quantities are linear combinations of the corresponding quantities in the two phases.
In particular, the free energy of a cell reads:
 \begin{equation}
{\cal F}_W =f {\cal F}^I + (1-f){\cal F}^{II} + {\cal F}_e  \label{totalfree_CPA} ,
\end{equation}
where $f=V^I/V_W$ is the volume fraction of the nucleus, ${\cal F}_e$ is given by Eq.~\eqref{free_e}, and ${\cal F}^{I(II)}$ is a short-hand notation for the baryonic free energy density of homogeneous matter (see Eq.~\eqref{free}) evaluated at densities $(\rho_p^{I(II)},\rho_n^{I(II)})$.

In the CLD model, the existence of an extended interface between the nucleus and the gas is taken into account by adding a surface and a Coulomb term in the energy density of each cell \citep{Avancini:2012bj}. The Coulomb and surface terms are given by:
\begin{eqnarray}
\mathcal{F}_{\rm Coul}&=&2 f e^2 \pi \Phi R^2 \left(\rho_p^I-\rho_p^{II}\right)^2 \ , \\
\mathcal{F}_{\rm surf}&=&3 f  \sigma / R  \ ,
\end{eqnarray}
where $R=(3V^I/4\pi)^{1/3}$ is the radius of the nucleus, $\sigma$ is the surface energy coefficient, and $\Phi$ is given by :
\begin{equation}
\Phi=\frac 1 5 \left ( 2+f-3f^{1/3}\right ) \ .
\end{equation}
For this work, we have used the functional form for the
surface energy coefficient $\sigma$ proposed in Ref.~\cite{Lattimer:1991nc} 
which was validated on extended Thomas-Fermi calculations, and only depends on the proton fraction of the nucleus $y_p^I=\rho_p^I/\rho^I$:
\begin{eqnarray} \label{eq:surf_tension}
\sigma&=&\sigma_0\frac{2^{p+1}+b_s}{{y_p^I}^{-p}+b_s+(1-y_p^I)^{-p}}h_t \ ,
\end{eqnarray}
with
\begin{eqnarray}
h_t&=&\left(1-(T/T_c)^2\right)^2 \ ,  \ {\rm{if}} \, T < T_c \nonumber \\
h_t&=&0 \, , \  {\rm {if}} \, T \geq T_c \ , \nonumber \\
T_c&=&87.76\left(\frac{%K_{\rm sat}
K_0}{375}\right)^{1/2}\left(\frac{0.155}{\rho_0}\right)^{1/3}y_p^I(1-y_p^I) \ . \label{eq:Tc}
\end{eqnarray}
In this expression, $\sigma_0$, $b_s$ and $p$ are free parameters that must be optimized for each nuclear model. 
To do so, we consider the prediction of the CLD model for the mass of a nucleus of mass number $A$ and proton number $Z$ in the vacuum:
\begin{eqnarray}
    M(A,Z)&=&m_p Z+ m_n (A-Z) \nonumber \\
    &+& \frac{A}{\rho_0(y_p)}  
    \epsilon(\rho_0(y_p),\delta,T=0) \nonumber \\
    &+& 4\pi R^2 \sigma+\frac 3 5 \frac{Z^2}{R} \ ,
    \label{eq:mass} 
\end{eqnarray}
where $y_p=Z/A$, $\delta = 1- 2y_p$, the nuclear radius is $R=(4\pi\rho_{0}(y_p)/3)^{-1/3} A^{1/3}$, and the bulk density $\rho_{0}(y_p)$ is given by the
equilibrium density of nuclear matter at proton fraction $y_p$, defined by 
$\partial\epsilon/\partial\rho|_{y_p,\rho_{0}(y_p),T=0}=0$.
For each model, the associated surface parameters $\sigma_0$, $b_s$ and $p$ are determined by a $\chi^2$-fit of Eq.~\eqref{eq:mass} to the experimental Atomic Mass Evaluation (AME) 2016 \citep{Audi:2017asy}.

Finally, since we are considering finite temperature matter, a translational free energy term should also be considered, which accounts for the droplet center-of-mass motion at temperatures beyond the crystallization temperature. 
If we consider the thermal motion as for an ideal gas, the translational term reads \citep{Haensel:2007yy}
\begin{eqnarray}
\mathcal{F}_{\rm trans} &=& \frac{F_{\rm trans}}{V_W}=
\frac{T}{V_W}\left[\ln\left(\frac{ \lambda^{3}}{V_W A_{\rm heavy}^{3/2}}\right)-1\right] \nonumber \\
&=&\frac{f T}{V^I}\left[\ln\left(\frac{f \lambda^3}{{V^I}^{5/2}{\rho^I}^{3/2}}\right)-1\right] \ ,\label{etrans}
\end{eqnarray}
where the mass number of the heavy cluster is given by Eq.~\eqref{eq:Aheavy}, 
and the nucleon thermal wavelength is given by $\lambda=\left(\frac{2\pi \hbar^2}{MT}\right)^{1/2}$. 
As discussed by many authors \citep{Lattimer:1991nc,Sedrakian:1996,Magierski:2004sua,Martin:2016,DinhThi:2023ioy}, the ideal gas approximation is not very realistic in the dense medium of the inner crust. 
Following Ref.~\cite{DinhThi:2023ioy}, we therefore consider an effective mass number of the heavy cluster, $A^*$, where the gas density is removed to account for the fact that the unbound nucleons do not participate to the translational motion of the ion:
\begin{eqnarray} \label{eq:mstar}
A^*&=&A_{\rm heavy}(1-\rho^{II}/\rho^I)=V^I(\rho^I-\rho^{II}) \ .
\end{eqnarray}
Moreover, the finite size of the nucleus implies that the free volume available for the ion motion is reduced with respect to the whole volume of the cell~\citep{Lattimer:1985zf}. The volume fraction $f$ (defined after Eq.~\eqref{totalfree_CPA}) should be replaced by $f^*=V^I/V_W^*$ with $V_W^*=4/3\pi (R_W-R)^{3}$, where $R_W$ is the radius of the Wigner-Seitz cell. This volume is related with the volume of the cluster by $V_W^*=V^I/f(1-f^{1/3})^3$. 
If we substitute these expressions in Eq.~\eqref{etrans}, the translational free energy density becomes:
\begin{eqnarray} \label{eq:etrans}
\mathcal{F}_{\rm trans}^*  &=&
\frac{F^*_{\rm trans}}{V_W}=
\frac{f T}{V^I}\left[\ln\left(\frac{\lambda^3}{V_W^*A^{*3/2}}\right)-1\right] \, \label{etrans*} \\
&=&\frac{f  T} {V^I}\left[\ln\left(\frac{f\lambda^3}{{V^I}^{5/2}(\rho^I-\rho^{II})^{3/2}(1-f^{1/3})^3} \right)-1\right] \nonumber
\end{eqnarray}
Note that in the applications shown in this paper, the proton gas is always negligible, and the results would not be changed by subtracting only the unbound neutrons in Eqs.~\eqref{eq:mstar}-\eqref{eq:etrans}.

By minimizing the sum $\mathcal{F}_{\rm surf}+\mathcal{F}_{\rm Coul} + \mathcal{F}_{\rm trans}$ with respect to the size $R$ of the droplet, one gets:
\begin{eqnarray}
\mathcal{F}_{\rm surf}&=& 2\mathcal{F}_{\rm Coul} - 3\mathcal{F}_{\rm trans}-\dfrac{15f T}{2 V^I} \ .
\end{eqnarray}
This expression holds if we consider either the effective or ideal gas expression for the translational energy term.
With the addition of the surface, Coulomb and translational term, the total free energy density in the cell Eq.~\eqref{totalfree_CPA} is modified to:
 \begin{eqnarray}
{\cal F}_W &=& f {\cal F}^I + (1-f){\cal F}^{II} + {\cal F}_e \nonumber \\ &+&\mathcal{F}_{\rm surf} + \mathcal{F}_{\rm Coul} + \mathcal{F}_{\rm trans} \ . \label{totalfree} 
\end{eqnarray}

\subsection{Inclusion of light clusters} \label{sec:light-clusters}
 
In this work, besides considering a heavy cluster, we also add  the different H and He isotopes to the system, namely deuterons d$\equiv ^2$H, tritons t$\equiv ^3$H, heliums h$\equiv ^3$He, $\alpha-$particles $\alpha\equiv ^4$He and $6$He$\equiv ^6$He, along the line of Ref.~\cite{Pais:2019shp}. 
These light clusters are considered as structureless point-like, with an effective mass that depends on the density, and are added as new degrees of freedom to the system. In-medium effects are also taken into account in a two-fold way: via a parameter, $x_s$, that was fixed to experimental constraints, in the scalar cluster-meson coupling, and via a binding energy shift term, to take into account the Pauli principle, avoiding therefore double counting \citep{Pais:2019shp,Pais:2019jst}.

This amounts to adding to the Lagrangian density Eq.~\eqref{eq:lagrangian} the extra contribution of these new degrees of freedom:
\begin{equation} 
  \sum_{j=p,n} {\mathcal{L}_{j}} \,\, \rightarrow \sum_{j=p,n,{\rm d,t,h,\alpha,6He}}{\mathcal{L}_{j}}  .
\end{equation}
The Lagrangian density term for the fermionic clusters, t and h, has the same form as for the nucleons (see Eq.~\eqref{lagnucl}) with the substitution $g_v\rightarrow A_j g_v$, $g_s\rightarrow x_s A_j g_s$, $M^*\rightarrow M^*_j$.
Here, $A_j$ is the mass number of the composite particle, its effective mass is given by
\begin{eqnarray}
M_j^*&=&A_jM - x_sA_jg_s\phi_0 - \left(B_j^0 + \delta B_j\right) \, ,
\label{meffi2}
\end{eqnarray}
where $B_j^0$ is the isotope binding energy in the vacuum, and $\delta B_j$ 
and $x_s$ are effective modifications of the binding energy and  scalar meson coupling, accounting for the mass shift of the bound composite in the dense medium, see Ref.~\cite{Pais:2019shp} for details.
For the scalar-isoscalar $J=0$ particles ($\alpha$ and $^6$He) we have:
\begin{eqnarray}
{\mathcal{L}_j}&=&\frac{1}{2} (i D^{\mu}_{j} \phi_{j})^*
(i D_{\mu j} \phi_{j})-\frac{1}{2}\phi_{j}^* \pc{M_{j}^*}^2
\phi_{j},
\end{eqnarray}
with 
$iD^{\mu }_j = i \partial ^{\mu }-A_jg_{v} \omega^{\mu }$, while the Lagrangian density of the vector-isoscalar $J=1$ particle (deuteron) reads:
\begin{eqnarray}
\mathcal{L}_{j}&=&\frac{1}{4} (i D^{\mu}_{j} \phi^{\nu}_{j}-
i D^{\nu}_{j} \phi^{\mu}_{j})^*
(i D_{j\mu} \phi_{j\nu}-i D_{j\nu} \phi_{j\mu})\nonumber\\
&&-\frac{1}{2}\phi^{\mu *}_{j} \pc{M_{j}^*}^2 \phi_{j\mu} \ .
 \end{eqnarray}
The number densities of the different light composites are given by Eq.~\eqref{eq:density} as for the nucleon densities, but the distributions depend on the quantum statistics:
\begin{equation}
f_{0 j \pm}= \frac{1}{\eta+e^{(\epsilon_{0 j} \mp \nu_j)/T}} \, ,
\end{equation}
with $\nu_j=\mu_j - A_jg_v V_0^{(0)}$, and  $\eta=1\ (-1)$ for fermions (bosons). Note that at the temperatures considered in the present study the $\alpha$ particles do not condensate.

The contribution of the light clusters is added to the bulk free energy ${\cal F}$ and all other thermodynamic quantities defined in Eqs.~\eqref{energy}, \eqref{entropy}, \eqref{free}, \eqref{pressure}.
The chemical equilibrium condition with the nucleons implies the definition of the chemical potentials for the composite particles ($j=d,t,h,\alpha,^6$He):
\begin{equation}
\mu_{j}=N_j\mu_n+Z_j\mu_p  \, ,
\end{equation}
where $N_j(Z_j)$ is the neutron (proton) number of the composite.

The total baryon density and the condition of charge neutrality read
\begin{eqnarray}
\rho&=&f\rho^I+(1-f)\rho^{II}+ \sum_{j}A_{j}\rho_{j} \label{rho} \, ,\\
\rho_p&=&f\rho_p^I+(1-f)\rho_p^{II}+ \sum_{j}Z_{j}\rho_{j} \, , \label{rhop} \\ 
&=&\rho_e=y_p\rho \ ,
\end{eqnarray} 
with ${j={\rm d,t,h,\alpha,^6He}}$, and $y_p$ the global proton fraction. 
We note that Eq.~\eqref{rhop} holds because $\rho_{j}^I=0$.

Finally, we define the following quantities, as in Ref.~\cite{Avancini:2012bj}: the mass fraction of the gas (unbound nucleons) is written as:
\begin{equation}
Y_{\rm free}= (y_p^{II}+y_n^{II})(1-f)\rho^{II}/\rho \ , \\
\end{equation}
with $y_{p(n)}^{I(II)}=\rho_{p(n)}^{I(II)}/\rho^{I(II)}$. 
The total mass fraction of the light composites is given by
\begin{equation}
 Y_{\rm light}= \sum_{j={\rm d,t,h,\alpha,^6He}}y_{j}^{II}(1-f)\rho^{II}/\rho \, , \\   
\end{equation}
and the mass fraction of the heavy ion is given by
\begin{equation} \label{eq:yheavy}
Y_{\rm heavy}= f\rho^{I}/\rho \, ,
\end{equation}
with $y_j^{I(II)}$
the mass fraction bound in each of the light clusters
\begin{equation}    
y_{j}^{I(II)}=A_{j}\frac{\rho_{j}^{I(II)}}{\rho^{I(II)}} \ .
\end{equation}

\subsection{Equilibrium composition}

The fractions of the different components of the Wigner-Seitz cell in equilibrium are determined from the minimization of the total free energy \citep{Baym:1971ax,Lattimer:1985zf,Bao:2014,Pais15}, which in our model includes the light clusters as well as the surface, Coulomb, and  translational terms of the heavy cluster. This minimization is done with respect to four variables: the size of the nuclear droplet, $R$, its baryonic and proton density, $\rho^{I}$, $\rho_p^I$, and its volume fraction within the cell, $f$. The equilibrium conditions become:
\begin{eqnarray}
\mu_n^I&=&\mu_n^{II} +\frac{3 T}{2  V^I \rho^I}  
- \frac{\mathcal{F}_{\rm surf}}{f} \beta_p \ , \\
\mu_p^I&=&\mu_p^{II}-\frac{2\mathcal{F}_{\rm Coul}}{f(1-f)(\rho_p^I-\rho_p^{II})} \nonumber \\
 &+&\frac{3 T}{2  V^I \rho^I} + \frac{\mathcal{F}_{\rm surf}}{f} \beta_n  \ , \\
P^{I}&=&P^{II}-\frac{2\mathcal{F}_{\rm Coul}}{(\rho_p^I-\rho_p^{II})}\left(\frac{\rho_p^I}{f}+\frac{\rho_p^{II}}{(1-f)}\right) \nonumber \\
&+&\mathcal{F}_{\rm Coul}\left(\frac{3}{f}
+\frac{1}{\Phi}\frac{\partial\Phi}{\partial f}\right) +\frac{\mathcal{F}_{\rm surf}}{f}\left(\beta_n\rho_p^I-\beta_p\rho_n^I\right) \nonumber \\
&-&\frac{5  T}{V^I}
-\frac{2\mathcal{F}_{\rm trans}}{f} \ ,
\end{eqnarray}
with
\begin{eqnarray}
   \beta_n&=&\frac{\rho_n^I}{\rho^{{I}^2}}\left\{p\frac{(1-y_p^I)^{-1-p}-y_p^{{I}^{-1-p}}}{{y_p^I}^{-p}+b_s+(1-y_p^I)^{-p}} + \frac{1}{h_t}\frac{d h_t}{dy^I_p}\right\} \nonumber \, , \\
    \beta_p&=&\frac{\rho_p^I}{\rho^{{I}^2}} \left\{p\frac{(1-y_p^I)^{-1-p}-y_p^{{I}^{-1-p}}}{{y_p^I}^{-p}+b_s+(1-y_p^I)^{-p}} - \frac{1}{h_t}\frac{d h_t}{dy^I_p}\right\} \nonumber \, , \\
y_p^I&=&\frac{\rho_p^I}{\rho^I} \, .
\end{eqnarray}
If instead we consider the effective translational energy, we get
\begin{eqnarray}
\mu_n^I&=&\mu_n^{II} +\frac{3  T}{2  V^I (\rho^I-\rho^{II})(1-f)}  
- \frac{\mathcal{F}_{\rm surf}}{f} \beta_p   \, , \\
\mu_p^I&=&\mu_p^{II}-\frac{2\mathcal{F}_{\rm Coul}}{f(1-f)(\rho_p^I-\rho_p^{II})} +\frac{3 T}{2  V^I (\rho^I-\rho^{II})} \nonumber \\ &+& \frac{\mathcal{F}_{\rm surf}}{f} \beta_n   \, , \\
P^{I}&=&P^{II}-\frac{2\mathcal{F}_{\rm Coul}}{(\rho_p^I-\rho_p^{II})}\left(\frac{\rho_p^I}{f}+\frac{\rho_p^{II}}{(1-f)}\right) \nonumber \\
&+&\mathcal{F}_{\rm Coul}\left(\frac{3}{f}
+\frac{1}{\Phi}\frac{\partial\Phi}{\partial f}\right) +\frac{\mathcal{F}_{\rm surf}}{f}\left(\beta_n\rho_p^I-\beta_p\rho_n^I\right)  \nonumber \\
&+&\frac{3 f T}{2 V^I(\rho^I-\rho^{II})}\left(\frac{\rho^I}{f}+\frac{\rho^{II}}{1-f}\right) \nonumber \\
&-&\frac{ T}{V^I}
\left(\frac{15}{2}-\frac{1}{1-f^{1/3}}\right)-\frac{2\mathcal{F}_{\rm trans}^*}{f}
\end{eqnarray}

We note that there are additional terms in both the chemical potentials and the mechanical equilibrium conditions compared to those that would be obtained in a simple thermodynamic phase equilibrium $\mu^I=\mu^{II},P^I=P^{II}$. 
These terms result from the inclusion of the surface, Coulomb, and translational energy terms in the total energy minimization. 
The Coulomb repulsion and the translational energy induce additional positive terms, while the surface tension reduces the internal cluster pressure. Since we are considering a surface tension parameterization that depends on $y_p^I=\rho_p^I/\rho^I$, there is also an extra term that appears in the equilibrium conditions.

\subsection{Impurity parameter} \label{sec:qimp}

Let us now define the impurity factor, $Q_{\rm imp}$.
The number densities of the heavy ion and of the light composites are given by:
\begin{eqnarray}
 \rho_{\rm heavy}&=&\rho \frac{Y_{\rm heavy}}{A_{\rm heavy}}  =\frac{1}{V_W} \\
  \rho_{\rm light}&=&\sum_{j={\rm d,t,h,\alpha,^6He}} \rho_{j}  
\end{eqnarray}
with $Y_{\rm heavy}$ given by Eq.~\eqref{eq:yheavy}, and the number of nucleon in the heavy cluster being given by Eq.~\eqref{eq:Aheavy}. 
With these definitions, we can now write the number fraction of the heavy cluster and of the light clusters:
\begin{eqnarray}
 X_{\rm heavy}&=&\frac{\rho_{\rm heavy} }{\rho_{\rm heavy} + \rho_{\rm light} }  \, ,\\
X_{\rm j}&=& \frac{\rho_{\rm j} } { \rho_{\rm heavy} + \rho_{\rm light} } .
\end{eqnarray}
The average and square average atomic number of clusters in a single cell, whatever their size, can then be written as
\begin{eqnarray}
    \langle Z\rangle&=&Z_{\rm heavy} X_{\rm heavy} + \sum_{j={\rm d,t,h,\alpha,^6He}} Z_{j} X_{j} \, , \\
     \langle Z^2\rangle&=&Z_{\rm heavy}^2 X_{\rm heavy} + \sum_{j={\rm d,t,h,\alpha,^6He}} Z_{j}^2 X_{j}  \, , 
\end{eqnarray}
which allows us to write 
$Q_{\rm imp} = \langle Z^2 \rangle-\langle Z \rangle^2$, with $Z_{heavy}$ given by Eq.~(\ref{eq:Zheavy}).

\begin{table}[htb!]
   \caption{Optimized surface parameters, namely $\sigma_0$, $b_s$, and $p$,  for the models considered in this work, fitted to the AME2016 mass table \citep{Audi:2017asy}.} 
   \setlength{\tabcolsep}{4pt}
 \begin{center}
  \begin{tabular}{lccccc}
    \hline \hline
    &$\sigma_0$ &$b_s$& $p$ \\
    & (MeV/fm$^2$)&& \\
    \hline
    FSU   & 1.17734&  21.18512&  3.0 \\
    FSU2R   & 1.15978&  25.63181&  3.0 \\
    NL3$\omega\rho$   & 1.15068&  25.20164&  3.0 \\
    NLset21   & 1.08889&  37.73938&  3.0 \\
    DD-ME2&  1.09358 & 5.47648 & 2.42 \\
    DD2&  1.03886 & 21.67598 & 3 \\
    TW&  1.14452 & 19.79133 & 3 \\
    \hline  \hline
  \end{tabular}
   \end{center}
\label{table:surface_mass_fit}
\end{table}
\begin{table*}
\caption{Symmetric nuclear matter saturation properties for all models considered in this work: saturation density $\rho_0$ (fm$^{-3}$),  binding energy per nucleon $E/A$, incompressibility $K_0$, skewness $Q_0$, kurtosis $Z_0$, symmetry energy coefficient $J_{\rm sym,0}$, slope $L_{\rm sym,0}$, curvature $K_{\rm sym,0}$, skewness $Q_{\rm sym,0}$, and kurtosis $Z_{\rm sym,0}$ of the symmetry energy (all in MeV). } 
\label{tab:2}
\setlength{\tabcolsep}{3.0pt}
      \renewcommand{\arraystretch}{1.1}
\begin{center}      
\begin{tabular}{ccccccccccc}
\hline \hline
\multirow{2}{*}{model} & \multirow{2}{*}{$\rho_0$} & \multirow{2}{*}{$E/A$} & \multirow{2}{*}{$K_0$} & \multirow{2}{*}{$Q_0$} & \multirow{2}{*}{$Z_0$} &  \multirow{2}{*}{$J_{\rm sym,0}$} & \multirow{2}{*}{$L_{\rm sym,0}$} & \multirow{2}{*}{$K_{\rm sym,0}$} & \multirow{2}{*}{$Q_{\rm sym,0}$} & \multirow{2}{*}{$Z_{\rm sym,0}$} \\ 
                  &                           &                               &                        &                                           &                &                                  &                                  &                   &           &                                    \\ 
\hline                  
FSU   & 0.148  & -16.32 & 228 & -521 & 2815 & 32.5 & 60 & -51 & 426 & -6332 \\
FSU2R   & 0.1505  & -16.26 & 238 & 135 & 4982 & 30.7 & 47 & 56 & 190 & -10917 \\
NL3$\omega\rho$   & 0.148  & -16.24 & 270 & 198 & 9304 & 31.6 & 55 & -8.07 & 1396 & -13602 \\
NLset21                  & 0.156                     & -16.12                        & 216                    & -339          & 6785                         & 29                               & 42                               & 55                               & 1146      & 14120                              \\
DDME2   & 0.152  & -16.14 & 251 & 479 & 4448 & 32. & 51 & -87 & 776 & -7047 \\
DD2   & 0.149065  & -16.02 & 243 & 168 & 5233 & 31.7 & 55 & -93 & 598 & -5152 \\
TW   & 0.153  & -16.24 & 240 & -539 & 3749 & 32.5 & 55 & -124.7 & 538 & -3307 \\
\hline 
\hline
\end{tabular}
\end{center}
\end{table*}

\begin{figure}[!htbp]
    \centering
   \includegraphics[width=\linewidth]{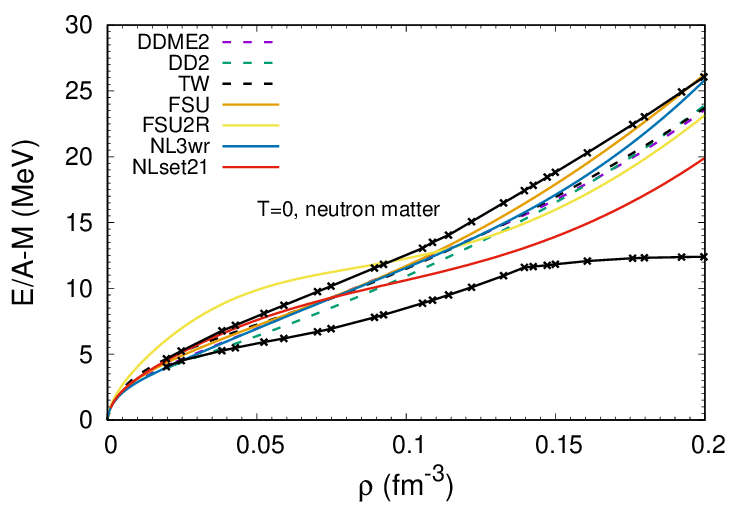} 
    \caption{Energy per baryon of pure neutron matter as a function of the total baryon density at $T = 0$ MeV for all the models considered in this work. The lines with stars indicate the overlapping region of several chiral-EFT calculations, taken from Ref.~\cite{Huth:2021bsp}. }
    \label{fig:EA_neut}
\end{figure}

\section{Results and Discussions}
\label{sec:results}

The calculations in this paper are performed using four non-linear models, FSU, FSU2R, NL3$\omega\rho$, NLset21, and three density dependent models, TW, DD2, and DDME2, to illustrate the model dependence of the results. 
These models fulfill several constraints, as discussed in Sect.~\ref{sec:method}.
As mentioned in Sect.~\ref{sec:light-clusters}, apart from the heavy cluster, we also include five light clusters, namely $^2$H, $^3$H, $^3$He, $^4$He and $^6$He. We perform the calculations at $\beta$-equilibrium, which is expected to be achieved in late cooling stages of proto-NS \citep{Camelio:2017nka}, and for temperatures close to the expected crystallization temperature of the NS crust. 
The optimal surface parameters obtained from the fit to the experimental mass table AME2016 for these models are displayed in Table~\ref{table:surface_mass_fit}.
Moreover, the symmetric nuclear matter parameters calculated at saturation density are listed in Table \ref{tab:2}. The models differ mainly in the density dependence of the symmetry energy.

To evaluate the reliability of the different models, we start by checking if the RMF models, non-linear and density dependent, are consistent with the latest ab-initio pure neutron matter calculations from chiral-EFT theory. 
To this aim, we plot in Fig.~\ref{fig:EA_neut}, the energy per neutron as a function of density.
We observe that only the FSU2R model fails this constraint, although the parametrization was constrained by the pure neutron-matter pressure from the ab-initio calculation obtained in Ref.~\cite{Hagen:2015yea}. 

\begin{figure}[!htbp]
    \centering
    \includegraphics[width=\linewidth]{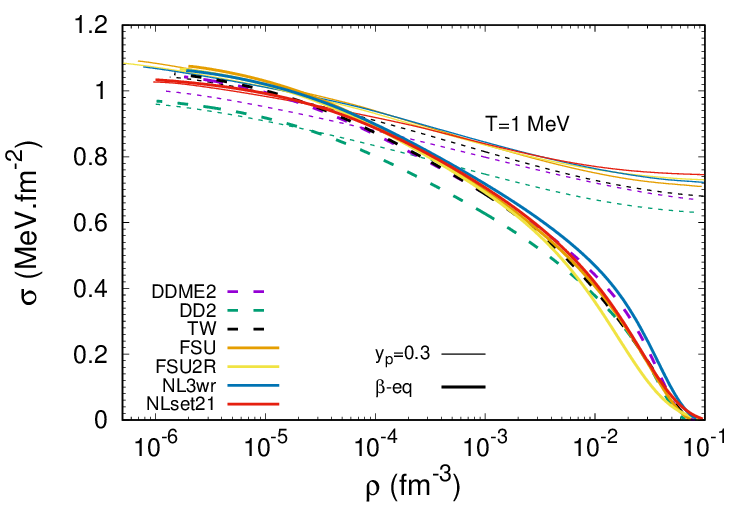} \\
    \includegraphics[width=\linewidth]{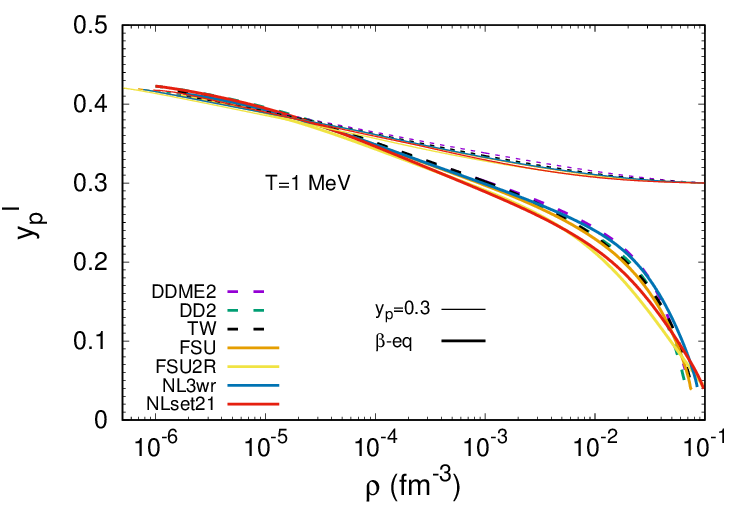}\\
    \includegraphics[width=\linewidth]{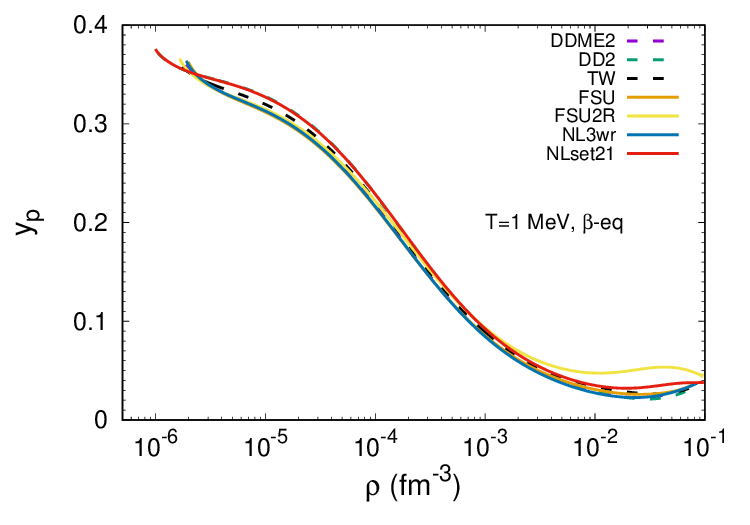}
    \caption{Cluster surface tension $\sigma$ (top panel), proton fraction in the dense phase $y_p^I$ (middle panel) and global proton fraction $y_p$ (bottom panel) in the inner crust  at $T=1$~MeV as a function of the total baryon density $\rho$ obtained for $\beta$-equilibrium  matter (thick lines)  and matter with fixed proton fraction $y_p = 0.3$ (thin lines in top and middle panels) for all the models considered in this work. 
    The results are obtained without the translational term, and the surface parameters are obtained from the fit to the mass table AME2016 \citep{Audi:2017asy}.}  
    \label{fig:surface-tension}
\end{figure}

In Fig.~\ref{fig:surface-tension}, we plot in the top panel the surface tension at $\beta$-equilibrium for the representative temperature $T=1$~MeV just slightly above the expected temperature of crystallization of the innermost part of the inner crust \citep{Carreau:2020gth,Thi:2021}, for the two classes of RMF models, non-linear (solid) and density-dependent (dashed). For comparison, the surface tension for a fixed proton fraction $y_p=0.3$, a value close to the neutron dripline, is also shown. The difference between the two results represents the effect on the surface energy of the density dependence of the isospin content of the heavy cluster in $\beta$ equilibrium.
In the middle panel, for the same temperature and models, we show the proton fraction in the dense phase against the baryonic density obtained for $\beta$-equilibrium matter (thick) and matter with fixed proton fraction (thin). The global proton fraction at $\beta$-equilibrium is plotted  as a function of the density in the bottom panel.
As one can see from Eq.~\eqref{eq:surf_tension}, the surface tension depends only on the proton fraction of the cluster $y_p^I$. 
The strong decrease of the surface energy with density thus reflects the increasing neutron richness of the equilibrium nucleus, which in turn is due to the decreasing global proton fraction as a function of density (see bottom panel). 
This general feature, as well as the disappearance of the surface tension at the crust--core transition, is common to all models. 
This behavior is in contrast to the case of a constant proton fraction. There, the proton fraction of the nucleus decreases very slowly (see middle panel), from $\sim 0.4$ in the very low density limit to the global value, 0.3, close to the dissolution of the clusters. This behavior is in turn reflected in the surface tension (see top panel).

\begin{figure}[!htbp]
    \centering
    \includegraphics[width=\linewidth]{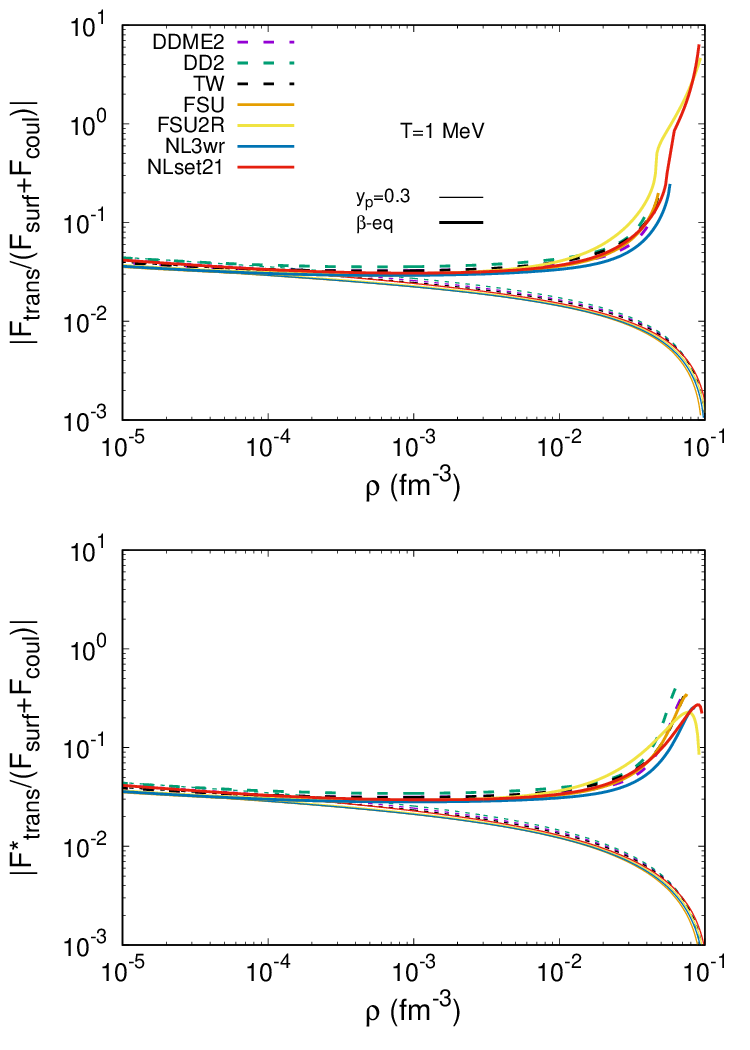}
    \caption{Absolute value of the translational free energy divided by the sum of the Coulomb and surface energies terms as a function of the total baryon density $\rho$ obtained in a CLD calculation with five light clusters at $T=1$~MeV in $\beta$-equilibrium matter (thick lines) and matter with fixed proton fraction $y_p = 0.3$ (thin lines) for all the models considered in this work. The top (bottom) panels are obtained with the translational term of ideal-gas, $F_{\rm trans}$ (with the corrections on the finite-size and in-medium effects, $F^{\star}_{\rm trans}$).}
    \label{fig:ratio}
\end{figure}

\begin{figure*}[!htbp]
    \begin{tabular}{c c}
    \includegraphics[width=0.5\linewidth]{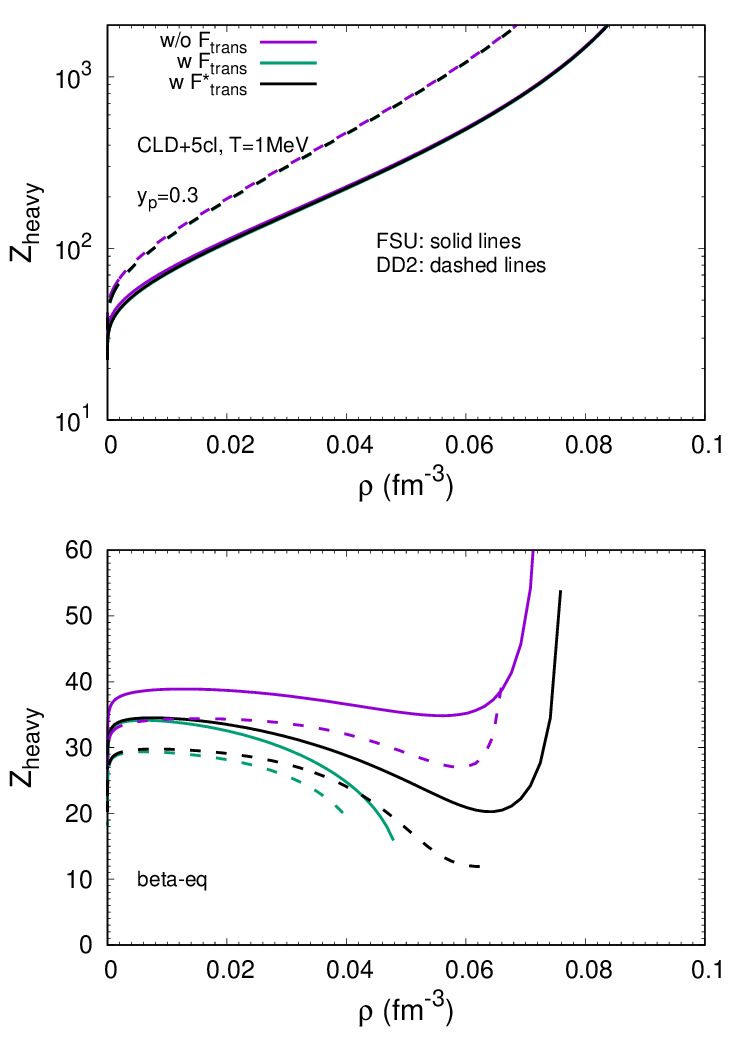} & \includegraphics[width=0.5\linewidth]{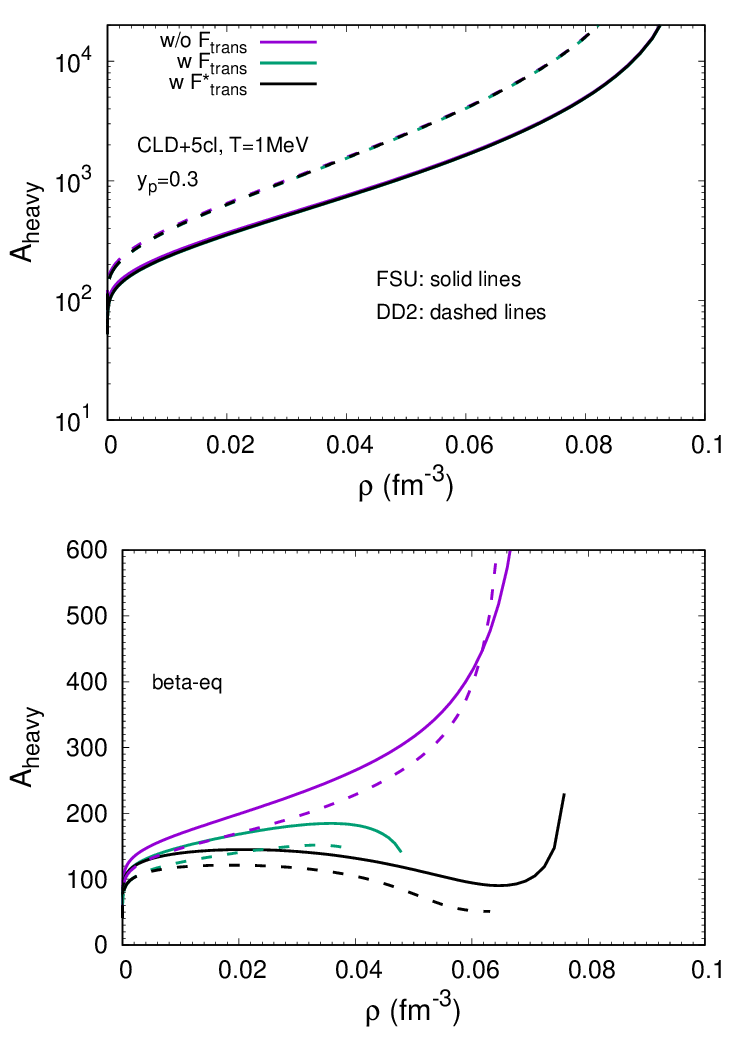}
    \end{tabular}
    \caption{Number of protons (left) and nucleons (right) in the heavy cluster as a function of the total baryon density $\rho$ in a CLD calculation with 5 light clusters for the FSU (solid lines) and DD2 (dashed lines) interactions with a fixed proton fraction of $y_p = 0.3$ (top panels) and $\beta$-equilibrium (bottom panels), at $T = 1$ MeV. The different colors represent the results obtained without the translational free energy (violet lines), with the ideal-gas translational term $F_{\rm trans}$ (green lines) and the corrected translational term $F^{\star}_{\rm trans}$ (black lines).} 
    \label{fig:Z-heavy}
\end{figure*}

In Fig.~\ref{fig:ratio}, we plot the absolute value of the ratio between the translational free energy term and the sum of the Coulomb and surface terms. 
Again, the specificity of the NS crust governed by $\beta$-equilibrium (thick lines) can be appreciated by the comparison with the fixed proton fraction case, also shown in the figure (thin lines).
In the $\beta$-equilibrium case, from the very low density limit up to $\sim 0.03$~fm$^{-3}$, the translational free energy term (with or without considering an effective mass number of the cluster) is much smaller than the sum of the Coulomb and surface terms. 
If the proton fraction of the droplet does not vary much with density, as is the case for $y_p=0.3$, the translational term in the inner crust becomes completely negligible, particularly for $\rho \gtrsim 0.01$ fm$^{-3}$, at least at the relatively low temperatures corresponding to crustal crystallization.
However, the effect is reversed in the innermost part of the inner crust when the $\beta$ equilibrium is considered, which is due to the strong decrease of the surface tension in the very neutron rich $\beta$ equilibrated matter observed in Fig.~\ref{fig:surface-tension}.
This is in line with the results shown in Ref.~\cite{DinhThi:2023ioy}, see their Fig.~6. 
As discussed at length in that work, in the global minimization of the cell free energy, the Coulomb plus surface term tend to favor the low-energy configurations as it is the case at $T=0$, corresponding to highly bound heavy clusters. Conversely,
the translational term tends to favor high-entropy configurations as in a classical gas, corresponding to small mass numbers for the droplet.   
The increase of the ratio between the two terms at high density means that the heavy clusters are going to melt in the dense medium in a much more effective way when this term is accounted for. 

This simplified picture occurs if we neglect the effect of the dripped neutrons on the translational motion of the droplet. With the inclusion of the effective mass number for the cluster, the translational degree of freedom is reduced in the dense medium which lowers this ratio and moderates the effect. Still, the effect of including the translational degrees of freedom in the variational equations is very important, as it can be seen from Fig.~\ref{fig:Z-heavy}.

Indeed, in this figure, we show how the number of protons (left) and nucleons (right) in the heavy cluster evolves with the density, fixing the temperature at $T=1$ MeV. 
We only consider the FSU and DD2 models, but the qualitative results are the same for all the models used in this work.  
We consider the three different cases: without (violet) and with the translational free energy, the latter considering both its ideal-gas version (green) and including the effective mass number of the cluster (black).
The composition of matter in the NS crust (lower panel, $\beta$ equilibrium) is very different from the case of dilute nuclear matter at constant proton fraction (upper panel).  
Looking at the top panels, the two models and the three different calculations behave similarly: the number of protons and of nucleons increase until the homogeneous matter is achieved. For the $\beta$-equilibrium calculation (bottom panels), the behavior is different: for both models, neglecting the translational term (violet lines) the composition of the liquid crust is very close to what is expected for the catalyzed ground-state composition, with an atomic number slowly varying according to the crustal depth, in the range $Z_{\rm heavy} \approx 30-40$ (left panel), and a faster increase for the nucleon number, in the range $A_{\rm heavy} \approx 100 - 400$ (right panel), and a sharp increase at the crust--core transition where matter becomes homogeneous. The difference between including or not the effective mass number in the translational term, i.e. the difference between considering $F_{\rm trans}^*$, Eq.~\eqref{etrans*}, instead of $F_{\rm trans}$, Eq.~\eqref{etrans}, is to make the decrease of the droplet proton and nucleon numbers smoother.  With the effective translational energy term (black lines), and for densities above $\sim 0.06$~fm$^{-3}$, the behavior of FSU is similar to the case without $F_{\rm trans}$: the number of protons and of nucleons increases at the crust--core transition where the clusters dissolve. The DD2 model follows a different trend and does not show a steep increase at the crust--core transition.
No heavy droplets survive at $T=1$~MeV above a density of $\rho\approx 0.04-0.05$~fm$^{-3}$ using $F_{\rm trans}$, as we can see from the bottom left panel (green lines).
This is because the effective number of nucleons has a strong correlation with the density of the  gas $\rho^{II}$ (recalling that $A^*=A_{\rm heavy}(1-\rho^{II}/\rho^I)=V^I(\rho^I-\rho^{II})$). 
In $\beta$ equilibrium, this quantity is quite high and close to the density of the cluster, when the baryonic density approaches the dissolution density of the clusters, making the cluster size to further decrease, $V^I=f V_W^*(1-f^{(1/3)})^{-3}$. This is in line with Ref.~\cite{DinhThi:2023ioy}, see their Figs.~7, 9 and 11. 

\begin{figure}[!htpb]
    \centering
    \includegraphics[width=\linewidth]{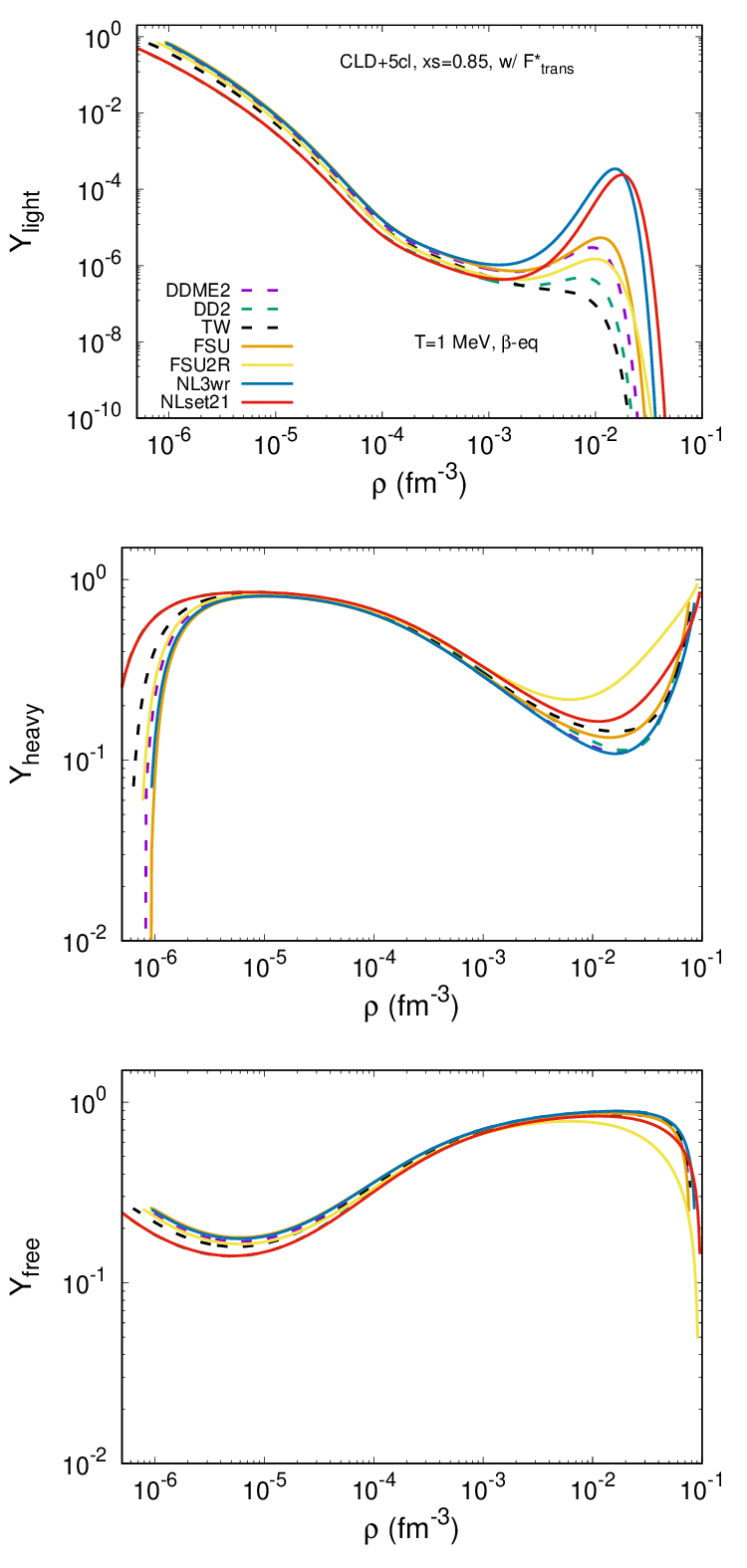}
    \caption{Mass fractions of the light clusters (top panel), heavy cluster (middle panel), and of the unbound nucleons (bottom panel), as a function of the total baryon density $\rho$ in the inner crust at $T=1$~MeV for all the models considered in this work. The results are obtained employing the corrected translational term $F^{\star}_{\rm trans}$.} 
    \label{fig:mass-fraction-beta}
\end{figure}

\begin{figure}[!htpb]
    \centering
    \includegraphics[width=\linewidth]{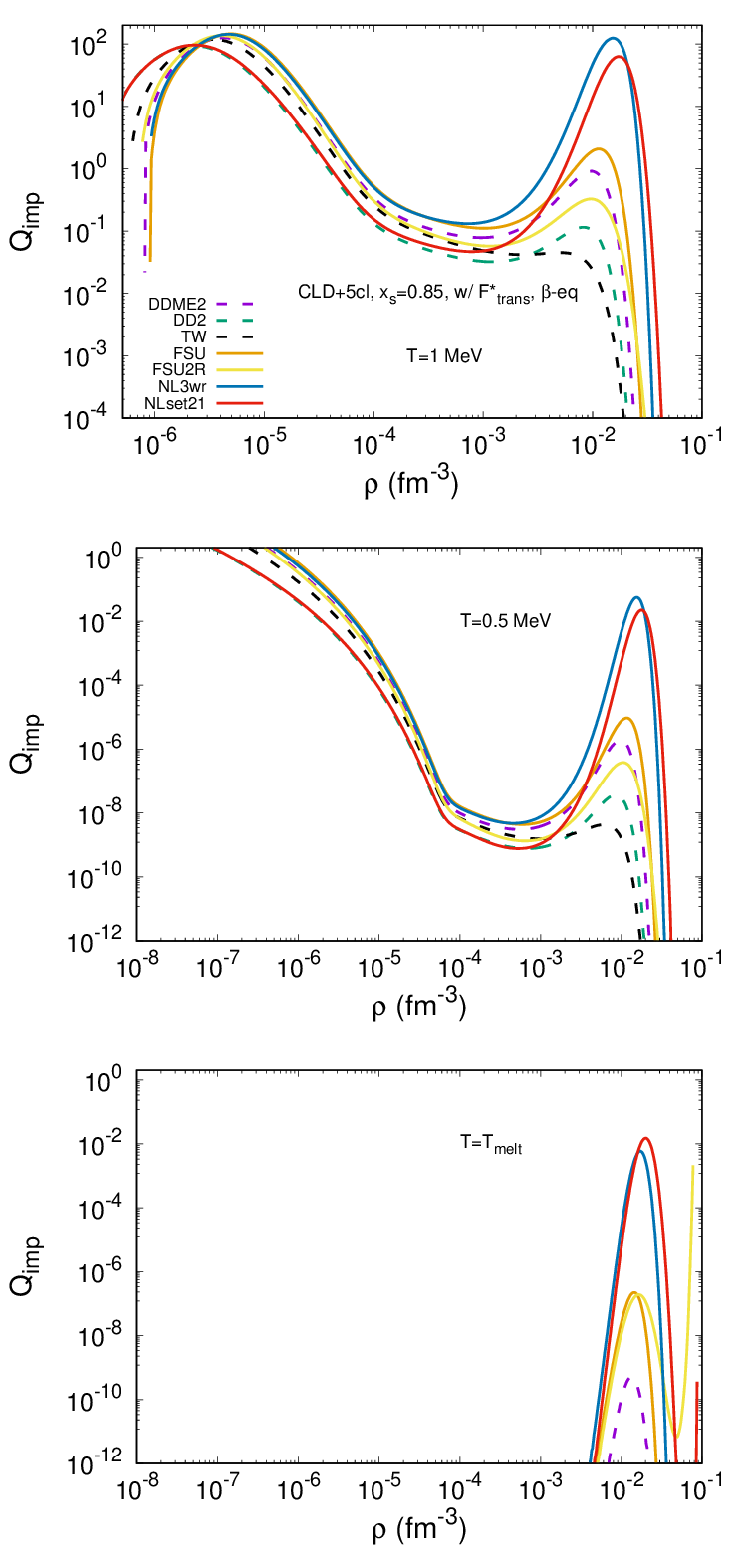}
    \caption{Impurity parameter $Q_{\rm imp}$ as a function of the total baryon density $\rho$ in a CLD calculation with 5 light clusters for all the models considered in this work in $\beta$-equilibrium, and at $T=1$ MeV (top panel), $T=0.5$ (middle panel), and at the crystallization temperature $T_{\rm melt}$ (bottom panel). Results are obtained employing the corrected translational term $F^{\star}_{\rm trans}$. }
    \label{fig:Qimp}
\end{figure}

In Fig.~\ref{fig:mass-fraction-beta}, we show the mass fractions as a function of the density for the light (top panel), and heavy (middle panel) clusters, together with the unbound nucleons (bottom panel). 
We only consider the most realistic prescription for the translational term  with the effective mass number of the cluster, Eq.~\eqref{etrans*}, and both non-linear (solid lines) and density-dependent (dashed lines) models at the representative temperature of $T=1$~MeV. 
The abundance of light clusters is systematically more important in the case of non-linear models, but it is globally small for all models, particularly in the inner crust region.  
This can be understood from the fact that the mass fraction of the heavy cluster dominates in most of the crust, implying that the volume fraction of dilute matter is too small to accommodate an important contribution of light composites together with unbound neutrons.  
Compared to Ref.~\cite{DinhThi:2023ioy}, where a full multi-component plasma calculation suggested a dominance of He and Li isotopes in the innermost part of the inner crust, we  observe that in this study these clusters were extremely neutron-rich, with mass numbers of the order of $A_{\rm heavy}\approx 20$. 

\begin{figure}[!htpb]
    \centering
    \includegraphics[width=\linewidth]{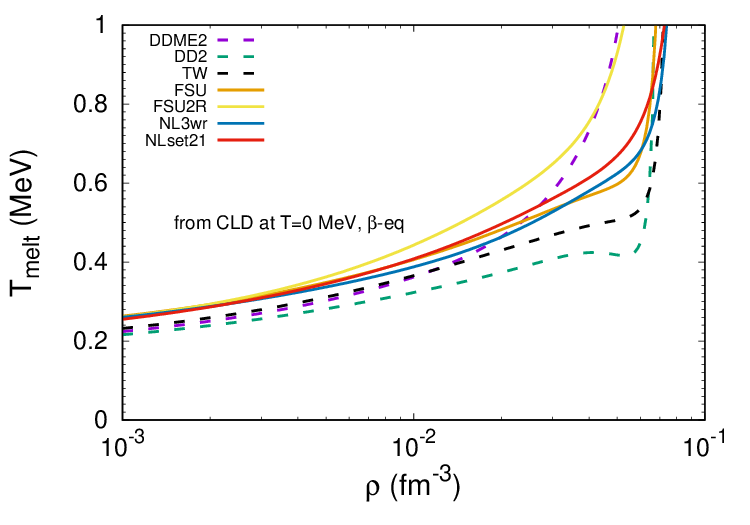}
    \caption{Crystallization temperature, $T_{\rm melt}$, as a function of the total baryon density $\rho$, with $Z_{\rm heavy}$ and $R_W$ obtained in a CLD calculation at $T=0$ and $\beta$-equilibrium for all the models considered in this work.} 
    \label{fig:Tmelt}
\end{figure}

\begin{figure}[!htpb]
    \centering
    \includegraphics[width=\linewidth]{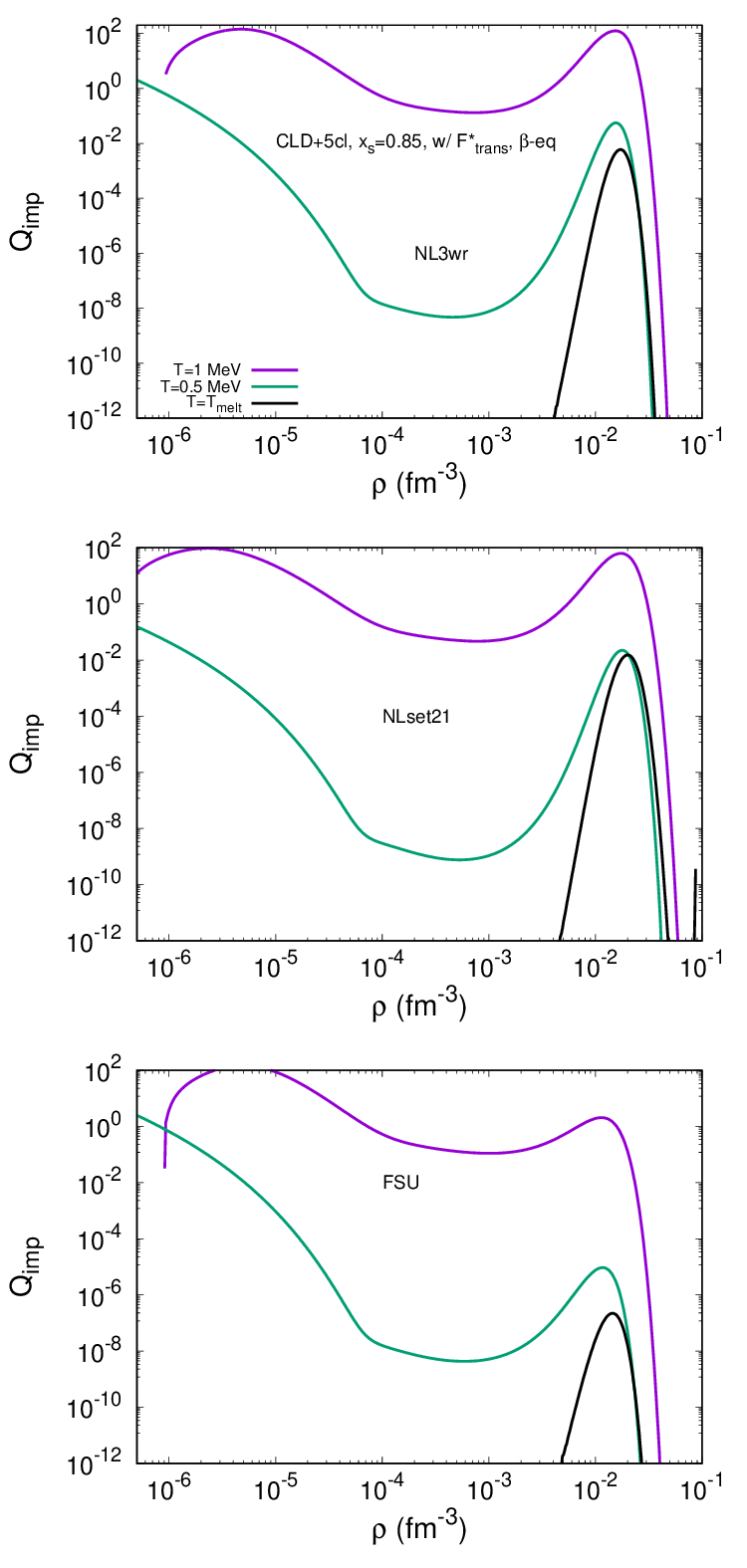}
    \caption{Impurity parameter $Q_{\rm imp}$ as a function of the total baryon density $\rho$ in a CLD calculation with 5 light clusters at $\beta$ equilibrium for $T=1$~MeV (dark blue lines), $T=0.5$~MeV (green lines) and $T=T_{\rm melt}$ (black lines), for NL3$\omega\rho$ (top panel), NLset21 (middle panel), and FSU (bottom panel). The results consider the corrected translational term $F^{\star}_{\rm trans}$.} 
    \label{fig:Qimp2}
\end{figure}

We turn now to the discussion of the impurity parameter.
In Fig.~\ref{fig:Qimp}, we show the impurity parameter calculated at $T=1$~MeV (top panel), 0.5~MeV (middle panel), and at the crystallization temperature $T=T_{\rm melt}$ (bottom panel), estimated as \citep{Haensel:2007yy} 
\begin{equation}
 T_{\rm melt} = \frac{Z_{\rm heavy}^2 e^2} {175\times R_{W}} \ ,
 \label{eq:tmelt}
\end{equation}
with $Z_{\rm heavy}$ and $R_{W}$
calculated from a CLD calculation at $T=0$ and in $\beta$ equilibrium. 
The crystallization temperature is plotted as a function of the density in Fig.~\ref{fig:Tmelt}. 
We notice that, as the density increases, the DD models present a lower impurity parameter as compared to the NL models. 
This is due to the fact that for the DD models the light-cluster mass fractions is lower, as can be seen in Fig.~\ref{fig:mass-fraction-beta}. 
Also, when we decrease the temperature to 0.5 MeV, the $Q_{\rm imp}$ values are much lower, especially for the DD models. 
The same happens when we calculate this parameter at the crystallization temperature: 
indeed, for DD2 and TW, this quantity is zero, since 
already at $T=0.5$~MeV and $\rho \gtrsim 0.01$~fm$^{-3}$, $Q_{\rm imp} \lesssim 10^{-8}$ (see middle panel of Fig.~\ref{fig:Qimp}), and $T_{\rm melt} \gtrsim  0.5$~MeV for $\rho > 0.04$~fm$^{-3}$ (see Fig.~\ref{fig:Tmelt}).
Similarly, for DDME2, $Q_{\rm imp}$ is also very low (notice the scale). 
On the other hand, for NL3$\omega\rho$ and NLset21, we have a peak of about $Q_{imp}\sim 10^{-2}$ at $\rho \sim 10^{-2}$~fm$^{-3}$.
We also notice that as the temperature decreases, the peak reached for each model decreases in magnitude but it happens at about the same density, as we can see from Fig.~\ref{fig:Qimp2}. 
This is very different from the results found in Ref.~\cite{DinhThi:2023hfp} (see their Fig.~15), where light H and He isotopes were not considered, but a full multi-component plasma calculation was performed accounting for fluctuations of the heavy cluster.
Indeed, in Ref.~\cite{DinhThi:2023hfp}, the change in temperature does not seem to particularly affect the amplitude of the maximum of $Q_{\rm imp}$, but it shifts the density at which the peak occurs.

\section{Conclusions} 
\label{sec:conclusions}

In this paper, we have calculated the abundance of light clusters (namely, the different isotopes of H and He) in thermodynamic conditions corresponding to the crystallization of the NS crust, using different RMF models and treating the light composites as independent structure-less quasi-particles. 
Such light clusters are expected to be formed in the dilute nucleon gas that characterizes the inner crust of the star while still in the liquid phase, and could therefore persist in the catalyzed crust due to the suppression of pycnonuclear reactions \citep{Chamel:2020pbw} below the crystallization temperature. 
The purpose of this study was to quantify the possible contribution due to such composite particles to the impurity factor that determines the resistivity of the NS crust. 

In all models, the contribution to the impurity factor that can be attributed to the presence of light cluster is negligible. 
This means that the nucleon fluid made of dripped neutrons and protons in the inner crust can be safely treated as an homogeneous fluid as in the mean-field approximation. 
As a consequence, if high $Q_{\rm imp}$ values exist in the inner crust as it is suggested by different astrophysical observations \citep{Pons:2013nea,Newton:2013,Deibel:2016vbc,Hambaryan:2017wvm,Tan:2018rhg}, its microscopic origin is likely to lie in the presence of either a deformed pasta mantle with defects \citep{Schneider:2014lia,Horowitz:2014xca,Caplan:2020ewl,Newton:2021vyd}, or of the fluctuations of the charge of the dominant (heavy) ion in each cell \citep{Carreau:2020gth,Fantina:2019lbd,DinhThi:2023ioy}, fluctuations that were not considered in this work.

\section*{ACKNOWLEDGMENTS}

This work was partially supported by national funds from FCT (Fundação para a Ciência e a Tecnologia, I.P, Portugal) under projects UIDB/04564/2020 and UIDP/04564/2020, with DOI identifiers 10.54499/UIDB/04564/2020 and 10.54499/UIDP/04564/2020, respectively, and the project 2022.06460.PTDC with the associated DOI identifier 10.54499/2022.06460.PTDC. H.P. acknowledges the grant 2022.03966.CEECIND (FCT, Portugal) with DOI identifier 10.54499/2022.03966.CEECIND/CP1714/CT0004. F.G., A.F. and H.D.T. ackowledge the support by the IN2P3 Master Project NewMAC and MAC, the ANR project `Gravitational waves from hot neutron stars and properties of ultra-dense matter' (GW-HNS, ANR-22-CE31-0001-01), the CNRS International Research Project (IRP) `Origine des \'el\'ements lourds dans l'univers: Astres Compacts et Nucl\'eosynth\`ese (ACNu)'.

\bibliographystyle{aa}

\end{document}